\documentclass{spie}
            \usepackage{graphicx}
            \usepackage{amsmath,amsfonts,amssymb}
            
\begin{document}
            \title{Effects of number scaling on entangled states in quantum mechanics}
            \author{Paul Benioff,\\
            Physics Division, Argonne National
            Laboratory,\\ Argonne, IL 60439, USA}
            \authorinfo{E-mail:pbenioff@anl.gov}

            \maketitle

            \begin{abstract}
            A  summary of number structure scaling is followed by a description of the effects of number scaling in nonrelativistic quantum mechanics. The description extends earlier work to include the effects on  the states of two or more interacting particles. Emphasis is placed on the effects on entangled states. The resulting scaling field is generalized to describe the effects on these states.  It is also seen that one can use fiber bundles with fibers associated with single locations of the underlying space to describe the effects of scaling on arbitrary numbers of particles.
            \end{abstract}


            \section{Introduction}
            In earlier work \cite{BenINTECH,BenNOVA,BenQSMF} the effect of scaling of number systems on some quantities in physics and geometry was described. The effect appears in the form of a  scalar scaling field that  appears in any physical quantity described as an integral or derivative over space and or time. In quantum mechanics this includes the momentum operator as a space derivative, $i\hbar d/dx$ and wave functions of particles.

            The work is based on the description of mathematical systems of different types as structures  and relations between the structures \cite{Shapiro,Barwise,Keisler}. Structures consist of a base set, a few basic operations and relations, and none or a few constants. They satisfy axioms relevant to the system type being considered.  The existence of the scaling field is based on the observation that number structures of each type can be scaled by arbitrary scaling factors. As a result the elements  of the base set of a number structure of a given type, (natural Integer, rational, real complex numbers) by themselves have no value.  They acquires value only as numbers in a structure.  Their value depends on the scaling of the structure.

            The scaling used here is linear in that the numbers in a scaled structure are all multiplied by a single scaling factor.  Recently a more general type of number scaling has been described and applied to fractal quantities \cite{Czachor1,Czachor2}. Here linear scaling will continue to be used because it is clearer and easier to understand.

             Fiber bundles \cite{Husemoller} are used to describe the effects of the scaling field in quantum mechanics. The base space  or manifold, $M$,  of the fiber bundle is $3$ dimensional Euclidean space.  For nonrelativistic quantum mechanics the fiber contains structures for complex numbers and Hilbert spaces and a representation of $M$.  The scalar scaling field appears in the fiber bundle approach as a connection between the fibers at different locations in $M$.  It connects or relates the components of  fibers at different $M$ locations.  It plays an essential role in quantities described by  integrals or derivatives  on $M$.

             The use of fiber bundles to describe  nonrelativistic quantum mechanics is not new. \cite{Moylan,Sen,Bernstein,Iliev,Asorey}. The quantum descriptions include an introductory description \cite{Bernstein}, a detailed mathematical development \cite{Iliev} and descriptions in which symmetry groups and semigroups  serve as the base space \cite{Sen,Asorey}.

             So far the effect of the scaling field in quantum mechanics has been limited to descriptions for single particle wave functions\cite{BenENSQM}. Here this description is extended to quantum mechanical descriptions of multiparticle wave functions, particularly those entangled in space degrees of freedom. As will be seen, this requires an expansion of the description of the scaling field and a change in the used of fiber bundles.

             The plan of this work is to review number scaling.  this is done in the next section.  This is followed by a section giving basic details on fiber bundles and connections.  Next comes a description of single particle quantum mechanics.  This is followed by the new material on two and multiparticle quantum mechanics.

             \section{Number scaling}\label{NS}

             The scaling of numbers is a simple concept.  It is easiest to see for the natural numbers. One notes that the natural numbers, $0,1,2,\cdots$ can be characterized as a set $N$ that is discretely well ordered with no greatest element. The values of the first and second element in the well ordering are $0$ and $1$. Let $N_{2}$ be the set of every other element of $N$.  The first element of $N_{2}$ has value $0$ and the second element has value $1$.  However the second element in $N_{2}$ has value $2$ in $N.$ One sees that the value of the second element of $N$ depends on whether it is considered as an element of $N$ or of $N_{2}.$

             This argument extends to any pair, $N_{m},N_{n}$ of subsets consisting of every $mth$ and $nth$ element of $N$. Let $n$ be a factor of $m.$  Then $N_{m}$ is a subset of $N_{n}$.  Let $j$ be an element of $N_{m}$.  The value $j$ in $N_{m}$ is related to that in $N_{n}$ by \begin{equation}\label{mvmj} mv_{m}(j)=nv_{n}(j).\end{equation} Here $v_{n}$ and $v_{m}$ are value functions that assign well ordering values to numbers in $N_{n}$ and $N_{m}.$

                Eq. \ref{mvmj} shows that elements of $N_{m}$ do not have intrinsic values of their own.  Their values depend on the subset and the well ordering of the subset containing them. For natural numbers this is restricted to all and only those subsets, $N_{n}$ where $m$ is a factor of $n.$

                 These results also apply to the other types of numbers,  The scaling for rational, real, and complex numbers differs from that of the natural numbers  in that the base set of numbers is the same for any scaling factor.  This is a consequence of the fact that these numbers are closed under division.

             Since the main concern here is the effect of scaling on quantum mechanics, the number type is limited to complex numbers. A discussion of number scaling for complex numbers is followed by a short discussion of the effect of number scaling on Hilbert spaces.

              \subsection{Complex numbers}\label{CN}

              The usual structure for complex numbers is given as \begin{equation}\label{BC}\bar{C}=\{C,\pm,\times,\div, ^{*},0,1\}.\end{equation} Here $^{*}$ denotes complex conjugation.  The structure is required to satisfy the axioms requiring that $\bar{C}$ is an algebraically complete field of characteristic $0$ \cite{complex}.

             Let $c$ and $d$ be  complex scaling factors. Define $\bar{C}^{c}$ and $\bar{C}^{d}$ to be  complex number structures associated with $c$ and $d$. The structure components are given by \begin{equation}\label{BCcd}\begin{array}{c}\bar{C}^{c}=\{B, \pm_{c}\times_{c},\div_{c},^{*_{c}},0_{c},1_{c}\}\\\\\bar{C}^{d}=\{B,\pm_{d}\times_{d}, \div_{d},^{*_{d}},0_{d},1_{d}\}.\end{array}\end{equation}The common base set for both structures is denoted by $B$. The subscripts $c$ and $d$ indicate structure membership.   Thus $1_{c}$ and $1_{d}$ are numbers in $B$ that have value $1$ in $\bar{C}^{c}$ and $\bar{C}^{d}$.

             It is  useful to let $\bar{C}$ denote a single complex number value structure. Elements of the base set $C$ in $\bar{C}$ are number values. One can then define bijective isomorphic value maps, $v_{c}$ and $v_{d}$ from $\bar{C}^{c}$ and $\bar{C}^{d}$ to $\bar{C}.$ For any number $b$ in $B$ $v_{c}(b)$ and $v_{d}(b)$ are the values $b$ has in $\bar{C}$. $v_{c}(b)$ and $v_{d}(b)$ can also be interpreted as the values $b$ has in $\bar{C}^{c}$ and $\bar{C}^{d}.$  The properties of value maps are given by \begin{equation}\label{vcst}\begin{array}{c} v_{c}(s\pm_{c}t) =v_{c}(s)\pm v_{c}(t),\;\; v_{c}(s\times_{c}t)=v_{c}(s)\times v_{c}(t),\\\\v_{c} (s\div_{c}t) =v_{c}(s)\div v_{c}(t),\;\; v_{c}(t^{*_{c}})=(v_{c}(t))^{*}.\end{array}\end{equation}

             Numbers in $B$ can also be described using the inverses of the value maps.   For example, $1_{c}=v^{-1}_{c}(1)$ and $1_{d}=v^{-1}_{d}(1).$ In general any number written  with a  scale factor subscript, as in $a_{d},$ has the value $a$ in $\bar{C}^{d}.$ If $c\neq d$ then the numbers $a_{d}$ and $a_{c}$ are different.  This is the case even though they have the same value in $\bar{C}$.

             The relation between the values, in $\bar{C}^{c}$ and $\bar{C}^{d},$ of a number in $B$ is given by an equation similar to Eq. \ref{mvmj}. One has \begin{equation}\label{dvvd1c} dv_{d}(b)= cv_{c}(b).\end{equation}  The relationship between the numbers in $B$ that have the same value in $\bar{C}^{c}$ and $\bar{C}^{d}$ is given by \begin{equation}\label{vdbfc}v_{d}(b)=\frac{c}{d}v_{c}(b)=v_{c} (\frac{c_{c}}{d_{c}}b).\end{equation} Here $c_{c}$ and $d_{c}$ are the numbers in $B$ with $c$ values, $c$ and $d$ in $\bar{C}.$ This equation shows the effect of moving scaling factors, as number values outside the value function,  to numbers inside the value function. Eq. \ref{vdbfc} can also be written in the equivalent form, \begin{equation}\label{vcbfd}v_{c}(b)=\frac{d}{c}v_{d}(b)=v_{d}(\frac{d_{d}}{c_{d}}b).
             \end{equation}

             These equations show clearly that the elements of the base set, by themselves, have no value.  They acquire value only as elements of the base set in a structure. The value of an element depends on the scaling factor of the structure containing it. There is also a separation between operations on numbers and operations on number values.  For example the number multiplications and divisions in the term, $(c_{c}/d_{c})b$, are in $\bar{C}^{c}.$ The multiplication and division in $(c/d)v_{c}(b)$ are in $\bar{C}.$

             Some words about the nomenclature used here and in the following are in order.  Structures will always be designated by an overline as in $\bar{C}^{c}$.  Base sets are without an overline as in $B.$\footnote{The letter $B$ is used instead of $C$ to distinguish it from the base set of $\bar{C}.$} Also elements of base sets of complex or any other number type, are referred to as numbers.  This distinguishes numbers from number values.  These are the values numbers have in a structure.

             Two representation of base set numbers are used.  One is in the form of a single letter as $b$ is used in Eqs. \ref{vdbfc} and \ref{vcbfd}.  The other is as a subscripted letter or number as in $a_{d}$.  $a_{d}$ is the base set number that has value $a$ in $\bar{C}^{d}$.  $1_{c}$ and $1_{d}$ are the numbers with value $1$ in $\bar{C}^{c}$ and $\bar{C}^{d}.$ But  $1_{c}$ is a different number from $1_{d}$.

             The relation between $a_{d}$ and $a_{c}$ is obtained from \begin{equation}\label{vdad} v_{d}(a_{d})= v_{c}(a_{c})=a\end{equation} and Eqs. \ref{vdbfc} and \ref{vcbfd}. One has\begin{equation} \label{adc}v_{d}(a_{d})=\frac{c}{d}v_{c}(a_{d})=v_{c}(\frac{c_{c}} {d_{c}}a_{d})=v_{c}(a_{c})= \frac{d}{c}v_{d}(a_{c})=v_{d}(\frac{d_{d}}{c_{d}}a_{c}). \end{equation} From this equation one gets \begin{equation}\label{vcad}v_{c}(a_{d})= \frac{d}{c}v_{c}(a_{c})=v_{c} (\frac{d_{c}} {c_{c}}a_{c}).\end{equation} Since $v_{c}$ is an isomorphism, \begin{equation}\label{adeq}a_{d}=\frac{d_{c}} {c_{c}}a_{c}.\end{equation}

             Note that the scaled structure $\bar{C}^{1}$ is distinct from $\bar{C}.$ If scaling is suppressed, then $\bar{C}^{1},$ as the only scaled structure, can be identified with $\bar{C}.$ This is the usual treatment of numbers in mathematics and physics.

             Let $s$ and $t$ be two numbers in $B$.  If both are in the same scaled structure, such as $\bar{C}^{c},$ then implementing the four field operations, and complex conjugation, on  the two numbers is well defined. Suppose, though, that $s$ is a number in $\bar{C}^{c}$ and $t$ is a number in $\bar{C}^{d}.$ Field operations are not defined for these numbers as the operations are defined only within structures, not between structures. One can define the operations on the values of the numbers. If $Op$ denotes any of the four field operations, then $v_{c}(s)Op v_{d}(t)$ is the resulting value of implementing $Op$ on $v_{c}(s)$ and $v_{d}(t).$ This is a different value from $v_{c}(s)Op v_{c}(t)$ as\begin{equation}\label{vcOp}v_{c}(s)Op v_{c}(t)= v_{c}(s)Op(\frac{d}{c})v_{d}(t).\end{equation}

             Maps that take the components of one scaled structure into those of another play an important role in what follows.  There are two types of maps.  The usual type is a value preserving number changing map of $\bar{C}^{d}$ onto $\bar{C}^{c}$  This map is defined by $a_{d}\rightarrow a_{c}$ and $Op_{d}=Op_{c}.$ Axiom validity is preserved under this map, $\bar{C}^{d}\rightarrow \bar{C}^{c}.$

             The other type of map is a number preserving, value changing map.  This is the map type studied here.   The effect of the map is to represent operations in one structure in terms of those in another structure in such a way that axiom validity is preserved.

              Let $Z^{d}_{c}$ be such a number preserving isomorphic map of the components of $\bar{C}^{d}$ onto those of $\bar{C}^{c}$.  It is defined by \begin{equation}\label{Zdc}\begin{array}{c}Z^{d}_{c}(s)= s, \;\;Z^{d}_{c}(\pm_{d})=\pm_{c},\;\; Z^{d}_{c}(\times_{d})= (\frac{\mbox{\small$c_{c}$}} {\mbox{\small$d_{c}$}} \times_{c}),\\\\ Z^{d}_{c}(\div_{d})=(\frac{\mbox{\small$d_{c}$}}{\mbox{\small$c_{c}$}}\div_{c}),\;\; Z^{d}_{c}(s^{*_{d}}) =\frac{\mbox{\small$d_{c}$}}{\mbox{\small$c_{c}$}}(s^{*_{c}}).\end{array}\end{equation} The multiplications and divisions implied in this definition are all in $\bar{C}^{c}.$ The representation of base set numbers as $d_{c}$ and $c_{c}$ means that these are base set complex numbers with values $d$ and $c$ in $\bar{C}^{c}.$ The scaling of the multiplication and division operations is needed to preserve axiom validity.

              The definition of $Z^{d}_{c}$ can be used to define a new structure, $\bar{C}^{d}_{c},$ that gives the components of $\bar{C}^{d}$ in terms of those of $C^{c}.$  The new structure is
              \begin{equation}\label{ZdcbC} Z^{d}_{c}\bar{C}^{d} =\bar{C}^{d}_{c}\end{equation}where \begin{equation}\label{bCdc}\bar{C}^{d}_{c}=\{B,\pm_{c},(\frac{c_{c}}{d_{c}}\times_{c}), (\frac{d_{c}}{c_{c}} \div_{c}),\frac{d_{c}}{c_{c}}(s^{*_{c}}),0_{d}, 1_{d}\}=\{B,\pm_{c}, (\frac{c_{c}}{d_{c}}\times_{c}),(\frac{d_{c}}{c_{c}} \div_{c}),\frac{d_{c}}{c_{c}}(s^{*_{c}}), \frac{d_{c}}{c_{c}}0_{c},\frac{d_{c}}{c_{c}}1_{c}\}.\end{equation}The validity of the complex number axioms for $\bar{C}^{d}_{c}$ follows from Eqs. \ref{vcbfd} and \ref{Zdc}.

              Let $s$ and $t$ be two numbers in $B$, Then \begin{equation}\label{Zdcaxval}
             \begin{array}{c}v_{d}(s\pm_{d}t)=v_{d}(s)v_{d}(+_{d})v_{d}(t)=v_{c}( \frac{\mbox{\small$d_{c}$}} {\mbox{\small$c_{c}$}}Z^{d}_{c}(s))v_{c}(Z^{d}_{c}(\pm_{d}))v_{c}(\frac{\mbox{\small $d_{c}$}} {\mbox{\small$c_{c}$}}Z^{d}_{c}(t)) =\frac{\mbox{\small$d$}}{\mbox{\small$c$}}v_{c}(s)\pm \frac{\mbox{\small$d$}}{\mbox{\small$c$}}v_{c}(t))=\frac{\mbox{\small$d$}}{\mbox{\small$c$}}v_{c} (s\pm_{c}t),\\\\v_{d}(s\times_{d}t)=v_{d}(s)v_{d}(\times_{d}) v_{d}(t)=v_{c}(\frac{\mbox{\small $d_{c}$}} {\mbox{\small$c_{c}$}}Z^{d}_{c}(s))v_{c}(Z^{d}_{c}(\times_{d}))v_{c}(\frac{\mbox{\small $d_{c}$}}{\mbox{\small$c_{c}$}}Z^{d}_{c}(t)) =\frac{\mbox{\small$d$}}{\mbox{\small$c$}}v_{c}(s) (\frac{\mbox{\small$c$}}{\mbox{\small$d$}}\times) \frac{\mbox{\small$d$}}{\mbox{\small$c$}} v_{c}(t))=\frac{\mbox{\small$d$}}{\mbox{\small$c$}} v_{c}(s\times_{c} t)\\\\v_{d} (s\div_{d}t)=v_{d}(s)v_{d}(\div_{d}) v_{d}(t)=v_{c}(\frac{\mbox{\small$d_{c}$}} {\mbox{\small$c_{c}$}}Z^{d}_{c}(s))v_{c}(Z^{d}_{c}(\div_{d}))v_{c}(\frac{\mbox{\small $d_{c}$}}{\mbox{\small$c_{c}$}}Z^{d}_{c}(t)) =\frac{\mbox{\small$d$}}{\mbox{\small$c$}}v_{c}(s) (\frac{\mbox{\small$d$}}{\mbox{\small$c$}}\div) \frac{\mbox{\small$d$}}{\mbox{\small$c$}} v_{c}(t))=\frac{\mbox{\small$d$}}{\mbox{\small$c$}} v_{c}(s\div_{c} t).\end{array}
             \end{equation}

             The fact that the values of the basic terms all scale by the same factor means that all equations are preserved under the map from $\bar{C}^{d}$ to $\bar{C}^{d}_{c}$.  This occurs because the terms on both sides of an equation are multiplied by the same scaling factor.   The map $Z^{d}_{c}$ and the relations among the value maps show that with one exception that values of all numbers are scaled.  The one exception is the number $0.$ The value of this number remains the same under all scalings.  This is the only number which can be identified with its value.  In a sense it acts like the number vacuum.

             \subsection{Hilbert spaces}\label{HS}

             Hilbert spaces as vector spaces over the complex number field are affected by complex number scaling. In general the setup is similar to that for the complex numbers.  The space, $\bar{H}$, where
             \begin{equation}\label{bH}\bar{H}=\{H,\pm,\cdot,\langle -,-\rangle,\psi\},\end{equation}represents a vector value structure for Hilbert spaces. $H$ is a set of vector values, $\pm,\cdot,$ and $\langle-,-\rangle$ denote linear superposition, scalar vector multiplication and a scalar product. Also $\psi_{c}$ and $\psi_{d}$ denote vectors with the value $\psi$ in $\bar{H}$. Here $\psi$ is an arbitrary vector value. The dimension of $\bar{H}$ is arbitrary, finite or denumerable. The structure, $\bar{C}$ is the scalar field for $\bar{H}$.

             The Hilbert space structures associated with $\bar{C}^{c}$ and $\bar{C}^{d}$ as scalars are given by \begin{equation}\label{bHc}\bar{H}^{c}=\{H_{c},\pm_{c},\cdot_{c},\langle -,-\rangle_{c},\psi_{c}\}, \end{equation} and \begin{equation}\label{bHd}\bar{H}^{d}=\{H_{d},\pm_{d},\cdot_{d},\langle -,-\rangle_{d}, \psi_{d}\}.\end{equation}As was the case for the complex numbers the base vector sets, $H_{c}$ and $H_{d}$ are the same, as $H_{c}=H_{d}.$  $\bar{H}^{c},\bar{H}^{d},\bar{H}$ are all required to satisfy the relevant axioms for Hilbert spaces.

             Value maps can also be defined for Hilbert spaces. They are similar to those for the complex numbers. One has for any vector $\phi$ in $H_{c}=H_{d}$, \begin{equation}\label{cvph}cv_{c}(\phi) =dv_{d}(\phi).\end{equation}From Eq. \ref{vdbfc} one has \begin{equation}\label{vdphi}v_{d}(\phi) =\frac{c}{d}v_{c}(\phi) =v_{c}(\frac{c_{c}}{d_{c}}\phi).\end{equation}The  vectors $\psi_{c}$ and $\psi_{d}$ in $H_{c}=H_{d}$ satisfy \begin{equation}\label{vcpsic}v_{d}(\psi_{d})= v_{c}(\psi_{c})=\psi.\end{equation}The vectors $\psi_{c}$ and $\psi_{d}$ differ only by scaling factors.  This can be seen from \begin{equation}\label{vcpsid}v_{d}(\psi_{d}) =\frac{c}{d}v_{c}(\psi_{d})=v_{c}(\frac{c_{c}}{d_{c}}\psi_{d})=v_{c}(\psi_{c})= \frac{d}{c}v_{d}(\psi_{c})=v_{d}(\frac{d_{d}}{c_{d}}\psi_{c}).\end{equation} This equation gives \begin{equation}\label{vcpsd}v_{c}(\psi_{d})=\frac{d}{c}v_{c}(\psi_{c})=v_{c}(\frac{d_{c}} {c_{c}}\psi_{c}).\end{equation}Since the value functions are isomorphic maps, one has \begin{equation}\label{psidc}\psi_{d}=\frac{d_{c}} {c_{c}}\psi_{c}.\end{equation}

             These results show that vectors in $\bar{H}^{d}$ and $\bar{H}^{c}$ differ by scaling factors only.  The projection of the components of $\bar{H}^{d}$ onto those of $\bar{H}^{c}$ is similar to that for the complex numbers.  The vector structure $\bar{H}^{d}_{c}$ is given by \begin{equation}\label{bHdc}\bar{H}^{d}_{c} =\{H_{c},\pm_{c},(\frac{c_{c}}{d_{c}}\cdot_{c}),\frac{d_{c}}{c_{c}}\langle\phi,\rho\rangle_{c}, \frac{d_{c}}{c_{c}}\psi_{c}\}.\end{equation}Here $\phi$ and $\rho$ are arbitrary vectors in $H_{c}=H_{d}.$ Also $\psi_{d}$ is an arbitrary vector in $H_{d}$.  This shows that the projection of the components of $\bar{H}^{d}$ onto those of $\bar{H}^{c}$ multiplies a vector by the  number factor $d_{c}/c_{c}.$ The scalar product component, $\frac{d_{c}}{c_{c}}\langle\phi, \rho\rangle_{c},$ is the same number as is  $\langle\phi,\rho\rangle_{d}.$ Both have value $\langle\phi,\rho\rangle$ in $\bar{C}^{d}.$ The value of $\langle\phi,\rho\rangle_{d}$ in $\bar{C}^{c}$ is $(d/c)\langle\phi,\rho\rangle.$

             \section{Fiber bundles and connections}\label{FBC}
             \subsection{Fiber bundles}\label{FB}
             Fiber bundles play an important role in many areas of physics.  They have also been used to describe some aspects of quantum mechanics \cite{Moylan,Sen,Bernstein,Iliev,Asorey}. They will be used here to describe the effects of number scaling on quantum states.

             A fiber bundle consists of a triple, $E,\pi,M$ where $E$ is the total space, $M$ is the base space, and $\pi$ is the projection of $E$ onto $M$ \cite{Husemoller}. Often $M$ is taken to be a space, time or space time manifold. Since the discussion will be limited to nonrelativistic quantum mechanics, $M$ is taken to be $3$ dimensional Euclidean space.

             In this case the fiber bundle becomes a trivial or product bundle where $E=M\times F$.  Here $F$ is the fiber.  The inverse of $\pi$ defines the fiber at point $x$ of $M$ by the relation
             \begin{equation}\label{pim1}\pi^{-1}(x)=(x,F)=F_{x}.\end{equation}

             Fibers can contain many different types of mathematical systems. Here the fiber $F$ contains pairs of Hilbert spaces and complex number structures for all scaling factors and a chart representation of $M$ as $\phi(M)=\mathbb{R}^{3}.$  The resulting fiber bundle is defined by
             \begin{equation}\label{MFCHM}\mathfrak{CHM}^{\cup}=M\times\bigcup_{c}(\bar{C}^{c}\times \bar{H}^{c})\times\mathbb{R}^{3},\pi,M.\end{equation} The fiber at point $x$ of $M$ is given by
             \begin{equation}\label{pim1CHM}\pi^{-1}(x)=F_{x}=\bigcup_{c}(\bar{C}^{c}_{x}\times \bar{H}^{c}_{x})\times\mathbb{R}^{3}_{x}.\end{equation} The structures $\bar{C}^{c}$ and $\bar{H}^{c}$ are given by Eqs. \ref{BCcd} and \ref{bHc}. The subscript, $x$ denotes local representations of these structures at point $x$ of $M$.

            A structure group $\mathcal{G}$  is associated with the bundle.  For each nonzero complex number value $d$  the $\mathcal{G}$ group element, $W_{d}$, acts on the fiber according to
             \begin{equation}\label{WdCHc}W_{d}(\bar{C}^{c}\times\bar{H}^{c})=\bar{C}^{dc}\times\bar{H}^{dc}.
             \end{equation}  The group, $\mathcal{G}$,  is commutative and associative. It also acts freely and transitively on the fiber.

             The presence of the structure group  with these properties implies that $\mathfrak{CHM}^{\cup}$ is a principal fiber bundle \cite{Drechsler}.  Structures at all levels $c$ in the fiber, $F$, are equivalent.  No one is to be preferred over another.

             \subsection{Connections}\label{C}
              Fields in physics can be represented  as sections over the fiber bundle. If $\beta$ is a scalar section then for each $x$ $\beta(x)$ is a complex number in $\bar{C}^{c_{x}}_{x}.$ If $\beta$ is  a vector section, then $\beta(x)$ is a vector in $\bar{H}^{c_{x}}_{x}.$ If the level $c_{x}$ is independent of $x,$ then the section is a level section.

              Physical quantities such as integrals or derivatives of $\beta$ over $M$ are not defined.  The reason is that mathematical combinations of quantities in fibers at different locations or levels are not defined. This is remedied by the use of connections.  These  map the contents of a fiber at one point to those at another point.

              There are different ways to define connections. Here the definition begins with a complex scalar field, $g,$ over $M$. This field induces a structure-group field, $W_{g},$ over $M$ where for each $x$ in $M$ and scaling factor $c$, \begin{equation}\label{Wgx}W_{g}(x)(\bar{C}^{c}_{x}\times\bar{H}^{c}_{x})= \bar{C}^{g(x)c}_{x}\times\bar{H}^{g(x)c}_{x}.\end{equation}

              Let $\beta_{g,c}$ be a vector section over $M$ where for each $x$ $\beta_{g,c}(x)$ is a vector in $\bar{H}^{g(x)c}_{x}.$ If $\beta_{g,c}$  is a scalar section, then $\beta_{g,c}(x)$ is a scalar in $\bar{C}^{g(x)c}_{x}.$  Let $y$ and $x$ be two points on $M$.  The number or vector in $\bar{C}^{g(x)c}_{x}$ or $\bar{H}^{g(x)c}_{x}$ that corresponds to  $\beta_{g,c}(y)$ in  $\bar{C}^{g(y)c}_{y}$ or $\bar{H}^{g(y)c}_{y}$ is given by \begin{equation}\label{Cgxyb}
              C_{g}(x,y)\beta_{g,c}(y)=Z^{g(y)}_{g(x)}V(x,y)\beta_{g,c}(y)=Z^{g(y)}_{g(x)}\beta_{g,c}(y)_{x} =\frac{g(y)_{g(x)}}{g(x)_{g(x)}}\beta_{g,c}(y)_{x}.\end{equation}

              The operator $V(x,y)$ maps a complex number Hilbert space pair at any level in $F_{y}$ to a pair at the same level in $F_{x}$. That is \begin{equation}\label{Vxy}V(x,y)(\bar{C}^{g(y)c}_{y} \times\bar{H}^{g(y)c}_{y}) =\bar{C}^{g(y)c}_{x}\times\bar{H}^{g(y)c}_{x}.\end{equation}The operator $Z^{g(y)}_{g(x)}$ is an extension of the definition in Eq. \ref{Zdc}  to include Hilbert spaces as in\begin{equation}\label{Zgygx}Z^{g(y)}_{g(x)}(\bar{C}^{g(y)}_{x} \times\bar{H}^{g(y)}_{x})= \bar{C}^{g(y)}_{g(x),x}\times\bar{H}^{g(y)}_{g(x),x}. \end{equation}The pair, $\bar{C}^{g(y)}_{g(x),x}$ and $\bar{H}^{g(y)}_{g(x),x}$ are described by Eqs. \ref{bCdc} and \ref{bHdc}. Here and in the following $c$ will be set equal to $1$ as the results are independent of the value of $c.$

              \section{Single particle quantum mechanics}\label{SPQM}
              \subsection{Position space representation}\label{PSR}
              The usual position space expression for a single particle wave packet is \begin{equation}\label{psi}
              \psi=\int\psi(x)|x\rangle dx.\end{equation}Since the concepts of number and number value, and vector and vector value are conflated, there is no reason to distinguish them.  This is the usual approach.  This is a special case of the number scaling where $g$ is a constant field with value $1.$ One can also regard Eq. \ref{psi} as a global expression because the mathematics used to describe the integral is global.  It is not located anywhere in $M$.  It is outside of space and time.

              One can also use local mathematics to describe the wave packet.  This is done by use of the fiber bundle.  The wave packet in the fiber $F$ at level, $c$  is described by \begin{equation}\label{psic}\psi_{c}= \int\psi(x)_{c}|x\rangle_{c} dx\end{equation}where the integral is over all points in $\mathbb{R}^{3}.$ The complex number $\psi(x)_{c}$  and the vector, $|x\rangle_{c}$ have values $\psi(x)$ and $|x\rangle$ in $\bar{C}^{c}$ and $\bar{H}^{c}.$

             This is similar to a global representation of the vector, $\psi_{c}$ as there is no reference to mathematical systems at points of $M$.  The corresponding local representations are those in the fibers at each point of $M$.  The representation at point $x$ is given by\begin{equation}\label{psix}\psi_{c,x}= \int_{x}\psi(y)_{c,x}|y\rangle_{c,x} dy\end{equation} The pair $\psi_{c,x}(y)$ and $|y\rangle_{c,x}$ are numbers and vectors in $\bar{C}^{c}_{x}$ and $\bar{H}^{c}_{x}.$ The subscript $x$ on the integral indicates that the integral is over all points in $\mathbb{R}^{3}_{x}$.  The representations at different points of $M$ and different levels, $c$, are all equivalent.

             In the following $c$ will be set equal to $1.$ It  will be retained as a subscript where needed to distinguish numbers and vectors  from number values and vector values.

             There is another way to describe the wave packet in the fiber bundle.  To achieve this it is useful to express the wave packet, $\psi$ in the form as \begin{equation}\label{psis}
              \psi=\int\psi(x)|x\rangle dx=\int\lambda(x)dx.\end{equation} Here $\lambda$ is considered as a vector field over $M$. As such it is raised to be a section on the fiber bundle. In the absence of scaling  $\lambda(x)$ is a vector in $\bar{H}^{1}_{x}.$

              The integral in the definition in Eq. \ref{psis} is not defined as the integrands are in Hilbert spaces at different $M$ locations. This can be fixed by choice of a reference location, $x$ in $M$ and using the connection to parallel transform the integrand to a Hilbert space in $F_{x}$. With no scaling the connection $C(x,y)=V(x,y)$ as the $Z$ operator in Eq. \ref{Cgxyb} is the identity. In this case \begin{equation}\label{Cxyl} C(x,y)\lambda(y)=V(x,y)\lambda(y)= \lambda(y)_{x}.\end{equation} Here $\lambda(y)_{x}$ is the same vector in $\bar{H}^{1}_{x}$ as $\lambda(y)$ is in $\bar{H}^{1}{y}.$

              To complete the process one should describe the collection  of vectors $\lambda(y)_{x}$ by an equivalent local field over $\mathbb{R}^{3}_{x}.$ This corresponds to the lifting of the domain of $\lambda$ from $M$ to $\mathbb{R}^{3}_{x}$ and replacing the integration variable $y$ by its chart value $\phi_{x}(y)=z_{y}$ in $\mathbb{R}^{3}_{x}.$

              The resulting expression for  wave packet $\psi_{x}$ is well defined in the fiber at $x$ as
              \begin{equation}\label{psix1}\psi_{x}=\int_{x}\psi(z_{y})_{x}|z_{y}\rangle_{x}dz_{y}.\end{equation} This integral, as a vector in $\bar{H}^{1}_{x},$ has the same value as does the global expression of Eq. \ref{psi}.

              In the presence of scaling the values are not the same. For each $y$ in $M$, $\lambda_{g}(y)$ is a vector in $\bar{H}^{g(y)}_{y}.$ The connection is needed to parallel transform each $\lambda_{g}(y)$ to a vector in the reference Hilbert space, $\bar{H}^{g(x)}_{x}$.  the result is \begin{equation}\label{Cglgy}C_{g}(x,y)\lambda_{g}(y)=\frac{g(y)_{g(x)}}{g(x)_{g(x)}} \lambda_{g}(y)_{x}.\end{equation} Here $\lambda_{g}(y)$ is the same vector in $\bar{H}^{g(y)}_{y}$ as $\lambda(y)$ is in $\bar{H}^{1}_{y}$ and $g(y)_{g(x)}$ and $g(x)_{g(x)}$ are numbers in $\bar{C}^{g(x)}_{x}$ with values $g(y)$ and $g(x).$

              Defining $\psi_{g,x}$ as an integral over $\mathbb{R}^{3}_{x}$ requires lifting the domains of $g$ and $\lambda$ from $M$ to $\mathbb{R}^{3}_{x}$.  The resulting expression for $\psi_{g,x},$ as an integral in $F_{x}$ is \begin{equation}\label{psigx}\psi_{g,x}=\int_{x} \frac{g_{x}(z)_{g(x)}}{g_{x}(z_{x})_{g(x)}}\psi(z)_{g(x)}|z\rangle_{g(x)}dz.
              \end{equation}  The relation between $g_{x}$ and $g$ is given  by
              \begin{equation}\label{gxzy}g_{x}(z_{y})=g(y).\end{equation} Here $z_{y}$ is the chart value in $\mathbb{R}^{3}_{x}$ of $y$ in $M$.

              Since $g_{x}(z_{x})_{g(x)}$ is independent of the integration variable, it can be moved outside the integral to obtain,\begin{equation}\label{psigx1}\psi_{g,x}= \frac{1}{g_{x}(z_{x})_{g(x)}}\int_{x}g_{x}(z)_{g(x)}\psi_{g}(z)_{x}|z\rangle_{g(x)}dz.
              \end{equation}In general the presence of the scaling factor means that the vector value of $\psi_{g,x}$ is different from that in the absence of scaling.  However the two values coincide if the scaling factor $g$ is a constant independent of location.

              It is useful to express $g$ as the exponential of a complex scalar field as in \begin{equation} \label{gye} g(y)=e^{\gamma(y)}.\end{equation}  The effect of scaling can be described in terms of either the $g$ field or the $\gamma$ field.

               Use of Eq. \ref{gye} in Eq. \ref{psigx1} gives \begin{equation}\label{psigx2}\begin{array}{l} \psi_{g,x}= (e^{-\gamma_{x}(z_{x})})_{g(x)} \int_{x}(e^{\gamma_{x}(z)})_{g(x)}\psi(z)_{g(x)}|z \rangle_{g(x)}dz\\\\\hspace{1cm}= (e^{-\gamma_{x}(z_{x})})_{g(x)}\int_{x}dz |z\rangle_{g(x)}\langle z|(e^{\gamma_{x}})_{g(x)} \psi_{g(x)}\rangle_{g(x)}.\end{array} \end{equation} The subscript, $g(x),$ on the factors indicates that the factors are numbers or vectors in $\bar{C}^{g(x)}_{x}$ or $\bar{H}^{g(x)}_{x}$ in the fiber at $x.$

              \subsection{Momentum space  representation}\label{MSR}
              The above representation of scaled quantum states for single particles is a space representation.  It is a consequence of representing a quantum state as a section over a fiber bundle based on position space. As is well known quantum states of particles have either position or momentum representations. Both are equivalent.

              A possible way to represent this using fiber bundles is to expand the base space $M$ to be the product of three dimensional momentum space and position space as in $M=P\times X.$  The corresponding fiber bundle is an expansion of Eq. \ref{MFCHM} to \begin{equation}\label{MFCHPX} \mathfrak{CHPX}^{\cup}=(P\times X)\times\bigcup_{c}(\bar{C}^{c}\times \bar{H}^{c}) \times\mathbb{R}^{3}_{P}\times\mathbb{R}^{3}_{X},\pi,P\times X.\end{equation}Here $\mathbb{R}^{3}_{P}$ and $\mathbb{R}^{3}_{X}$ are chart representations of $P$ and $X.$

              Quantum states can have either position representations or momentum representations.  Unlike classical states they cannot have combined position and momentum representations. A quantum state amplitude cannot be a function of both position and momentum. It follows from this that quantum states as sections on a fiber bundle can be either position space sections or momentum space sections.  A quantum state cannot be both  a position and a momentum based section.

              This can be accommodated in the fiber bundle by replacing the full projection, $\pi$ by two marginal projections, $\pi_{X}$ and $\pi_{P}.$ $\pi_{X}$ does not act on $P$ and $\pi_{P}$ does not act on $X.$  The inverses of these projections define fibers at position and momentum space representations. One has \begin{equation}\label{pim1x}\pi_{X}^{-1}(x)=\bigcup_{c}(\bar{C}^{c}_{x} \times \bar{H}^{c}_{x}) \times\mathbb{R}^{3}_{x,P}\times\mathbb{R}^{3}_{x,X}=F_{x}\end{equation} and \begin{equation}\label{pim1p}\pi_{P}^{-1}(p)=\bigcup_{c}(\bar{C}^{c}_{p} \times \bar{H}^{c}_{p}) \times\mathbb{R}^{3}_{p,P}\times\mathbb{R}^{3}_{p,X}=F_{p}.\end{equation}

              Position state representations of quantum states can be implemented as sections on the fiber bundle, $\mathfrak{CHPX}^{\cup}$ by replacing $\pi$ by $\pi_{X}$.  This is what has been done so far.  Momentum state representations can be implemented as sections on the fiber bundle with $\pi_{P}$ replacing $\pi.$  However momentum state representations can also be described in the fibers $F_{x}$ just as position state representations can be described in the fibers $F_{p}.$  Here the discussion will be limited to describing momentum state representations in $F_{x}.$

              The momentum representation of the state $\psi_{g,x}$ as a vector in $\bar{H}^{g(x)}_{x}$ in $F_{x}$ is obtained in the usual way from Eq. \ref{psigx2}. One obtains \begin{equation}\label{psigp2}\begin{array}{l}\psi_{g,x}= (e^{-\gamma_{x}(z_{x})})_{g(x)}\int_{x}dp|p\rangle_{g(x)}\langle p|e^{\gamma_{x}}\psi\rangle_{g(x)}\\\\\hspace{1cm}= (e^{-\gamma_{x}(z_{x})})_{g(x)}\int_{x}\int_{x}dpdq|p\rangle_{g(x)}\langle p|e^{\gamma_{x}}|q\rangle_{g(x)}\langle q|\psi\rangle_{g(x)}.\end{array}\end{equation}

              The momentum amplitude, $\langle p| e^{\gamma_{x}}|q\rangle_{g(x)}$ can be expanded to
              \begin{equation}\label{lpxeg}\langle p| e^{\gamma_{x}}|q\rangle  = \int_{x}\langle p|z\rangle e^{\gamma_{x}(z)}\langle z|q\rangle dz = \int_{x}e^{\frac{\mbox{\small $i(p_{x}-q_{x})z$}} {\mbox{\small $\hbar$}}}e^{\gamma_{x}(z)} dz.\end{equation} The subscript $g(x)$ has been removed to indicate that the factors, with subscripts as complex numbers in $\bar{C}^{g(x)}_{x}$ or vectors in $\bar{H}^{g(x)}_{x}$, have, as values, the factors  without subscripts.  Eq. \ref{lpxeg} is an equation in quantity values in the fiber at $x.$  In the following the subscript $g(x)$ be suppressed for the same reason.

              These results show that  in the momentum representation the product of the scaling factor and the wave vector is present as a convolution integral over an intermediate momentum.  This quite different from the space representation in which the scaling factor and wave vector are present as a product.

              The effect of the momentum operator, $\tilde{p}_{x}$, on $\psi_{g,x}$ is expressed in the fiber $F_{x}.$ For the momentum representation of $\psi_{g,x}$ as in Eq. \ref{psigp2} one has
              \begin{equation}\label{tpxps}\tilde{p}_{x}\psi_{g,x}= (e^{-\gamma_{x}(z_{x})})\int_{x} p|p\rangle\langle p|e^{\gamma_{x}}\psi\rangle dp.\end{equation}

              The action of $\tilde{p}_{x}$ on the space representation of $\psi_{g,x}$ in the fiber at $x$ is  obtained from Eq. \ref{psigx2}. One has \begin{equation}\label{tppsigx}\begin{array}{l} \tilde{p}_{x} \psi_{g,x}=(i\hbar)(e^{-\gamma_{x}(z_{x})}) \int_{x}\frac{d}{dz} (e^{\gamma_{x}(z)}\psi(z)|z \rangle) dz\\\\\hspace{1cm} =i\hbar e^{-\gamma_{x}(z_{x})} \int_{x}e^{\gamma_{x}(z)} (\Gamma_{x}(z)+\frac{d} {dz})\psi(z)|z \rangle dz.\end{array} \end{equation}  Eq. \ref{tppsigx} shows that\begin{equation}\label{tpegxz}\tilde{p}_{x} (e^{\gamma_{x}(z)}\psi(z)|z \rangle)= e^{\gamma_{x}(z)}(\tilde{p}_{x}(z)+i\hbar\Gamma_{x} (z))\psi(z)|z \rangle.\end{equation} In these expressions $\Gamma_{x}$ is the gradient of $\gamma_{x}$ as in \begin{equation}\label{Gma} \Gamma_{x}(z)=\nabla_{z}\gamma_{x}(z). \end{equation}

              This result shows that the effect of the localization of a global expression of a wave packet as an integral of a section, in the presence of a varying scaling field, introduces a vector field  as a momentum component.  With no field $\Gamma$ present, the equation is the usual expression for action of the momentum operator on the wave packet expressed directly as an integral over $\mathbb{R}^{3}_{x}.$ Note that if $\gamma$ is a constant field, then the usual expression is recovered as $\Gamma=0.$

              These results can be used to determine the effect of scaling on the single particle Hamiltonian.  The position representation of the Hamiltonian as an operator on the states in $\bar{H}^{g(x)}_{x}$  in the fiber $F_{x}$ is given by \begin{equation}\label{Hxz}\tilde{H}_{x}(z)=\tilde{K}_{x}(z)+\tilde{V}_{x}(z)=\frac{\hbar^{2}} {2m}\sum_{j=1}^{3} \frac{d^{2}}{dz_{j}^{2}}+V_{x}(z).\end{equation}

              The kinetic energy operator acting on $\psi_{g,x}$ can be obtained from the observation that \begin{equation}\label{d2dzj2}\frac{d^{2}}{dz_{j}^{2}}e^{\gamma_{x}(z)}\psi_{x}(x)= \frac{d}{dz_{j}}e^{\gamma_{x}(z)} (\frac{d}{dz_{j}}+\Gamma_{x,j}(z))\psi(z)= e^{\gamma_{x}(z)} (\frac{d}{dz_{j}}+\Gamma_{x,j}(z))^{2} \psi(z).\end{equation}  If desired, the right hand term can be expanded to $$e^{\gamma_{x}(z)}(\frac{d^{2}} {dz_{j}^{2}}+\frac{d}{dz_{j}} (\Gamma_{x,j}(z))+2\Gamma_{x,j}(z)\frac{d}{dz_{j}}+\Gamma_{x,j}(z)^{2}) \psi(z).$$

               The action of the Hamiltonian on the scaled wave packet, $\psi_{g,x}$ at $z$ in $F_{x}$ is given by \begin{equation}\label{Hz}\tilde{H}_{x}(z)e^{\gamma_{x}(z)}\psi_{x}(z)= (\frac{\hbar^{2}}{2m}\sum_{j=1}^{3} \frac{d^{2}}{dz_{j}^{2}}+V_{x}(z))e^{\gamma_{x}(z)}\psi(z) =\frac{\hbar^{2}}{2m}e^{\gamma_{x}(z)}\sum_{j=1}^{3} (\frac{d}{dz_{j}}+\Gamma_{x,j} (z))^{2}\psi(z) +V_{x}(z)\psi(z).\end{equation} The factor $e^{-\gamma_{x}(z_{x})}$ has been left out because it is in effect a fiber normalizing constant and the Hamiltonian does not act on it.  The kinetic energy can also be expressed as $\tilde{K}_{x}=p_{g,x}^{2}/2m$ where \begin{equation} \label{pgx} p_{g,x}(z)=i\hbar\sum_{j=1}^{3}(\frac{d}{dz_{j}}+\Gamma_{j} (z))=i\hbar(\nabla_{z}+ \Gamma_{x}(z)).\end{equation}

               One notes that the eigenvalue equation for the Hamiltonian is given by \begin{equation} \label{tHx}\tilde{H}_{x}\psi_{x}=E\psi_{x}.\end{equation} The exponential scaling factors do not appear as they cancel from both sides of the equation. The effect of scaling is already included in the kinetic energy operator.

               \section{Two or more particle quantum mechanics}\label{TPQM}
               \subsection{Two or more noninteracting particles}
               Quantum mechanics of two or more particles requires changes  in the fiber bundle framework and in the effects of the scaling field.  For two independent particles that do not interact with one another one can start with two separate fiber bundles, one for each particle. Bundles for particles $1$ and $2$ would be \begin{equation}\label{MFK12}\mathfrak{CHR}_{1}=M,F_{1}, \pi_{1},M\mbox{ and }\mathfrak{CHR}_{2}=M,F_{2},\pi_{2},M.\end{equation}  In the absence of scaling $F_{1}=\bar{C}\times\bar{H}_{1}\times\mathbb{R}^{3}\times\mathbb{P}^{3}.$ $F_{2}$ differs only in replacing $\bar{H}_{1}$ with $\bar{H}_{2}.$

               A product of the bundles can be defined by \begin{equation}\label{MFCHR12} \mathfrak{CHR}_{12} =M,(F_{1}\times F_{2}),\pi_{1}\times\pi_{2},M.\end{equation} $\pi_{1}$ and $\pi_{2}$ are projection maps from $M,F_{1}$ to $M$ and $M,F_{2}$ to $M$,  The inverse maps are given by
               \begin{equation}\label{pim112}\begin{array}{c}\pi^{-1}_{1}(x)=\bar{C}_{x},\bar{H}_{1,x}, \mathbb{P}^{3}_{x},\mathbb{R}^{3}_{x}\\\\\pi^{-1}_{2}(y)=\bar{C}_{y},\bar{H}_{2,y}, \mathbb{P}^{3}_{y},\mathbb{R}^{3}_{y}.\end{array}\end{equation}

               One can define wave packet states  for particles $1$ and $2$ directly in fibers $F_{1,x}$ and $F_{2,y}$ by $\psi_{1,x}=\int_{x}\psi_{1}(z)|z\rangle dz$ and $\psi_{2,y}=\int_{y}\psi_{2}(u) |u\rangle du.$ Taking a direct product of these states is not meaningful because the states are in fibers at different locations. This can be remedied by restricting consideration to particle $1$ and $2$ fibers at the same location, $x.$ Then the direct product of $F_{1,x}$ and $F_{2,x}$ can be expressed as \begin{equation}\label{F12x}F_{1,x}\times F_{2,x}=F_{1,2.x}= \bar{C}_{x},\bar{H}_{1,x}\times\bar{H}_{2,x},\mathbb{P}^{3}_{x},\mathbb{R}^{3}_{x}.
               \end{equation} The product state $\psi_{1}\times\psi_{2}$ is a state in the product Hilbert space.

               The introduction of scaling follows that described for single particle states. The scaling field is the same for both particles.  It affects the state for each particle just as it does for the single particle state.

               This description extends to a descriptions of the quantum states of $n$ noninteracting particles. The product of two fiber bundles as in Eq. \ref{MFCHR12} is extended to an $n$ fold product as in \begin{equation}\label{MFCHR1n} \mathfrak{CHR}_{1,n} =M,(\times_{j=1}^{n}F_{j}), \times_{j=1}^{n}\pi_{j},M.\end{equation} The inverse projection map for particle $j$ is given by \begin{equation}\label{pim1j}\pi^{-1}_{j}(x_{j})=\bar{C}_{x_{j}},\bar{H}_{j,x_{j}}, \mathbb{P}^{3}_{x_{j}},\mathbb{R}^{3}_{x_{j}}\end{equation}

               The fiber associated with the n fold product of the inverses of the $\pi_{j}$ is a fiber associated with $n$ different locations in $M$ as in \begin{equation}\label{Fx1n}
               F_{x_{1,\cdots,n}}=\times_{j=1}^{n}F_{x_{j}}.\end{equation} As was the case for two particle states, one can define wave packet states for the $jth$ particle in $F_{x_{j}}$.  However one cannot take a direct product of these states for different $j$ because they are in fibers at different locations of $M$.

               This can be fixed by restricting the $x_{j}$ to be the same, as in $x_{j}=x$ for all $j$. Then the fibers all coalesce into a fiber at one point, $x$, as in\begin{equation}\label{Fxn} F_{x}=\bar{C}_{x}\times_{j=1}^{n} \bar{H}_{x,j} \times\mathbb{R}^{3}_{x}\times\mathbb{P}^{3}_{x}.\end{equation} The projection operator, $\pi$ projects onto single points of $M$ with $\pi^{-1}(x)=F_{x}.$
               \subsection{Two interacting particles}

               This  fiber bundle framework for a description of the quantum mechanics of two or more noninteracting free particles does not work for the description of states of two or more interacting particles or or those in entangled states.  A simple example of such a state is a  two particle Slater determinant state \begin{equation}\label{sldt}\psi_{1,2}(x,y)=\frac{1}{\sqrt{2}} (\psi_{1}(x)\psi_{2}(y)-\psi_{1}(y)\psi_{2}(x)).\end{equation} The two particle state vector for such entangled states is \begin{equation}\label{psi12}\psi_{1,2}=\int \psi_{1,2}(x,y) |x,y\rangle dxdy.\end{equation}

               The methods used  to include the effects of scaling begin with the representation of the integrand of Eq. \ref{psi12} as a vector field as in \begin{equation}\label{plam}
               \psi_{1,2}=\int\lambda_{1,2}(x,y)dxdy.\end{equation}The vector field can be described as a section on an appropriately defined fiber bundle. Connections are used to parallel transform the section values at different fiber locations to a common location.  Integrals of the transformed section are well defined with the fiber at the reference location.

               This can be achieved by defining a fiber bundle by \begin{equation}\label{mfr12}
               \mathfrak{CHPR}_{1,2}^{\cup}=M,F_{1,2},\pi_{1,2},M.\end{equation}The fiber components are
               \begin{equation}\label{F12}F_{1,2}=\bigcup_{c}(\bar{C}^{c}\times (\bar{H}_{1,2})^{c}, \mathbb{P}^{3},\mathbb{R}^{3}.\end{equation}The Hilbert space, $(\bar{H}_{1,2})^{c}$, is the tensor product, $(\bar{H}_{1})^{c}\bigotimes (\bar{H}_{2})^{c}$ of the two single particle spaces.

               The main difference between this bundle and that for single particle states is that the projection, $\pi_{1,2}$ acts on pairs of points of $M$. The inverse of $\pi_{1,2}$ defines  fibers associated with point pairs in $M$ as in\begin{equation}\label{pi12}
               \pi_{1,2}^{-1}(x,y)=F_{1,2,x,y}=\bigcup_{c}(\bar{C}^{c}_{x,y}\times (\bar{H}_{1,2})^{c}_{x,y}), \mathbb{P}^{3}_{x,y},\mathbb{R}^{3}_{x,y}.\end{equation}

               At each point pair, $u,v$ in $M$, the section  value, $\lambda_{1,2}(u,v)$  is a vector in $(\bar{H}_{1,2})^{c}_{u,v}$ at some level $c.$ The integration implied in Eq. \ref{plam} requires parallel transformation of $\lambda_{1,2}(u,v)$ at the different point pairs to a common reference pair, $x,y.$ This is done by use of a connection defined on a scalar  scaling field, $h,$ defined on point pairs of $M$.

               The action of the connection is given by \begin{equation}\label{Chxu}C_{h}(x,y;u,v) \lambda_{1,2} (u,v)=\frac{h(u,v)_{h(x,y)}}{h(x,y)_{h(x,y)}}\lambda_{1,2}(u,v)_{x,y}. \end{equation}Here $h(u,v)_{h(x,y)}$ and $h(x,y)_{h(x,y)}$ are complex numbers in the base set of $\bar{C}^{h(x,y)}_{x,y}$ that have values, $h(u,v)$ and $h(x,y)$ in $\bar{C}^{h(x,y)}_{x,y}$. Also $\lambda_{1,2}(u,v)_{x,y}$ is the same vector in $(\bar{H}_{1,2})^{h(x,y)}_{x,y}$ as $\lambda_{1,2}(u,v)$ is in $(\bar{H}_{1,2})^{h(u,v)}_{u,v}.$

               Use of this connection enables the integrals in Eqs. \ref{psi12} and \ref{plam} to be defined   when $\lambda_{1,2}$ is a section.  The result is\begin{equation}\label{pishxy} \psi_{h,x,y}= \int_{x,y}\frac{h(w,z)_{h(x,y)}} {h(w_{x},z_{y})_{h(x,y)}}\lambda_{1,2}(w,z)_{x,y}dwdz= \frac{1}{h(w_{x},z_{y})_{h(x,y)}} \int_{x,y}h(w,z)_{h(x,y)}\psi_{1,2}(w,z)|w,z\rangle dwdz.
               \end{equation} The integral is over all points ,$w,z$ in $\mathbb{R}^{3}_{x,y}.$ The factor $h(w_{x},z_{y})_{h(x,y)}$, is moved outside the integral as it is independent of $w,z.$ $w_{x}$ and $z_{y}$ are the points in $\mathbb{R}^{3}_{x,y}$ that correspond to $x,y$ in $M$.

               The scalar scaling function $h(u,v)$ is defined as the geometric average  of $g(u)$ and $g(v)$ as in \begin{equation}\label{huvs} h(u,v)=\sqrt{g(u)g(v)}=e^{\rho(u,v)}\end{equation} where
               \begin{equation}\label{rhouv}\rho(u,v)=\frac{\gamma(u)+\gamma(v)}{2}.\end{equation} Use of this to replace $h$ in Eq. \ref{pishxy} gives \begin{equation}\label{psihxy}(\psi_{1,2})_{h,x,y}=
               (\psi_{1,2})_{g,x,y} =e^{-\rho_{x,y}(w_{x},z_{y})}\int_{x,y} e^{\rho_{x,y}(w,z)}\psi(w,z)|w,z\rangle dwdz \end{equation} as the final result. Here $\rho_{x,y}$ is the lifting of $\rho$ to a scalar field with domain $\mathbb{R}^{3}_{x,y}$ in the fiber at $x,y$.

               The action of the two particle momentum operator on $\psi_{1,2}$ is \begin{equation} \label{tp12}\begin{array}{l}(\tilde{p}_{1}+\tilde{p}_{2})(\psi_{1,2})_{g,x,y}= e^{-\rho_{x,y}(w_{x},z_{y})} i\hbar\int_{x,y} (\nabla_{w}+\nabla_{z})e^{\rho_{x,y}(w,z)} \psi(w,z)|w,z\rangle dwdz\\\\\hspace{1cm}= e^{-\rho_{x,y}(w_{x},z_{y})}i\hbar\int_{x,y} e^{\rho_{x,y}(w,z)} (\nabla_{w}+\nabla_{z}+ \frac{1}{2}(\Gamma(w)+\Gamma(z)))\psi(w,z) |w,z\rangle dwdz.\end{array}\end{equation}Here $\Gamma$ is the gradient of $\gamma,$ lifted into the fiber at $x,y.$ From this one has\begin{equation}\label{pt12} (\tilde{p}_{1}+\tilde{p}_{2})e^{\rho(w,z)}=e^{\rho(w,z)} i\hbar(\nabla_{w} +\frac{1}{2}\Gamma(w)+\nabla_{z}+\frac{1}{2}\Gamma(z)).\end{equation}

               As was the case for the single particle Hamiltonian, the effect of scaling on the two particle Hamiltonian shows up in the kinetic energy.  The kinetic energy component in the two particle Hamiltonian, $\tilde{H}_{1,2}$ in the fiber at $x,y$ is \begin{equation}\label{K12}
               K_{1,2}(w,z)=\frac{\hbar^{2}}{2m_{1}}(\nabla_{w}+\frac{1}{2}\Gamma(w))^{2}+ \frac{\hbar^{2}}{2m_{2}}(\nabla_{z}+\frac{1}{2}\Gamma(z))^{2}.\end{equation}This assumes particles $1$ and $2$ have masses $m_{1}$ and $m_{2}.$

               \subsection{Quantum mechanics for one or two particles}\label{QMOTP}

               So far the quantum mechanics for one particle and for two particles are based on different fiber bundles. For one particle  the fibers are associated with one point of $M.$ For two particles the fibers are associated with point pairs of $M$. It would be good to find a common bundle representation for both one and two particle quantum mechanics.

               One way to achieve this is to recognize that the choice of reference point pairs for the two particle case is arbitrary.  Any pair, $x,y$  will  suffice.  This suggests that one choose $x,y$ to be $x,x.$  In this case the contents of the fibers $F_{x,x}$ are given by \begin{equation}\label{Fxx}F_{x,x}= \bigcup_{c}(\bar{C}^{c}_{x,x}\times(\bar{H}_{1,2})^{c}_{x,x}), \mathbb{P}^{3}_{x,x}, \mathbb{R}^{3}_{x,x}. \end{equation}The contents of $F_{x,x}$ are identical to those in the fiber $F_{x}.$  This can be seen from \begin{equation}\label{bCxx}\begin{array}{c} \bar{C}^{c}_{x,x}\times (\bar{H}_{1,2})^{c}_{x,x}= \bar{C}^{c}_{x}\times(\bar{H}_{1,2})^{c}_{x} \\\\\mathbb{P}^{3}_{x,x} =\mathbb{P}^{3}_{x} \\\\\mathbb{R}^{3}_{x,x}=\mathbb{R}^{3}_{x}. \end{array}\end{equation}

               The description of the effect of the connection and the scaling field for the two particles is that given by Eq.\ref{Chxu} and following equations with $y=x.$ Eq. \ref{psihxy} becomes \begin{equation}\label{psihxx}(\psi_{1,2})_{h,x,x}=(\psi_{1,2})_{g,x,x} =e^{-\rho_{x,x}(w_{x},w_{x})}\int_{x,x} e^{\rho_{x,x}(w,z)}\psi(w,z)|w,z\rangle dwdz. \end{equation} This equation is equivalent to \begin{equation}\label{psihx} (\psi_{1,2})_{h,x}=(\psi_{1,2})_{g,x}=e^{-\gamma_{x}(w_{x})}\int_{x} e^{\rho_{x}(w,z)}\psi(w,z) |w,z\rangle dwdz.\end{equation}

               This  is an example of the observation that  quantum mechanics for two interacting particles can be included in fibers associated with single points of $M$.  The fiber in the bundle, $\mathfrak{CHM}^{\cup}$ of Eq. \ref{MFCHM} is expanded to\begin{equation}\label{F}F=\bigcup_{c} (\bar{C}^{c}\times \bar{H}_{1}^{c}+H_{1,2}^{c})\times\mathbb{R}^{3}.\end{equation}The fiber at point $x$ is\footnote{If desired, a representation of the momentum space as $\mathbb{P}^{3}_{x}$ can be added to the fiber at $x.$} \begin{equation}\label{pim1x12} \pi^{-1}(x)=(\bar{C}^{c}_{x}\times\bar{H}_{1,x}^{c}+H_{1,2,x}^{c})\times\mathbb{R}^{3}_{x}.
               \end{equation}The scaling field in the fiber is different for the single particle space than it is for the two particle space. For the one particle space it is given by $e^{\gamma_{x}(w)}.$ for the two particle space it is given by $e^{\rho_{x}(w,z)}.$

               \subsection{$n$ interacting particles}\label{QMNIP}

               The description for states of two entangled particles can be extended to describe the effects of scaling field on the states and quantum mechanics of $n$ interacting particles. The fiber locations are expanded from two locations  to $n$ locations as in $F_{x_{1},\cdots,x_{n}}.$  The Hilbert space in this fiber, is the $n$ fold tensor product of the single particle Hilbert spaces. Eq. \ref{psihxy} becomes\begin {equation} \label{psihx1n}(\psi_{1,n})_{h,x_{1,n}}=(\psi_{1,\cdots,n})_{g,x_{1,n}} =e^{-\rho_{x_{1,n}}(w_{x_{1}},\cdots, w_{x_{n}})}\int_{x_{1,n}} e^{\rho_{x_{1,n}}(w_{1,n})} \psi(w_{1,n})|w_{1,n}\rangle dw_{1},\cdots dw_{n} \end{equation}In this equation $x_{1,n}$ and $w_{1,n}$ denote $x_{1},\cdots,x_{n}$ and $w_{1},\cdots,w_{n}.$

               The only difference from the two particle entangled state description is that the scaling field $h(x,y)$ must be generalized to \begin{equation}\label{h1n}h(x_{1},\cdots,x_{n})=(\times_{j=1}^{n} g(x_{j}))^{1/n}=e^{\sum_{j=1}^{n}\gamma(x_{j})/n}=e^{\rho(x_{1,n})}.\end{equation}The subscript, $x_{1,n}$ on $\rho$ in the above equation indicates the lifting of $\rho$ from a function on $n$ tuples of points in $M$ to $n$ tuples of points in $\mathbb{R}^{3}_{x_{1,n}}.$
               The momentum and kinetic energy for the $jth$ particle in the fiber, $F_{x_{1,n}}$ are given by \begin{equation}\label{momj}\tilde{p}_{j}(w_{j})=i\hbar(\nabla_{w_{j}}+\frac{\Gamma (w_{j})}{n}) \end{equation} and \begin{equation}\label{Kj}K_{j}=\frac{\hbar^{2}}{2m_{j}} (\nabla_{w_{j}} +\frac{\Gamma (w_{j})}{n})^{2}.\end{equation}Here $m_{j}$ is the mass of the $jth$ particle.

               \subsection{Quantum mechanics for one  or more particles.}\label{QM12P}
               Inclusion of the quantum mechanics for $n$ interacting particles is included in the fiber bundle for one particle in the same way as was done for two particles.  The reference fiber location, $x_{1},\cdots,x_{n}$, in the $n$ particle fiber bundle, is chosen to be $x_{j}=x$ for $1\geq j\geq n.$ In this case $\bar{C}^{c}_{x,\cdots,x}=\bar{C}^{c}_{x}$ and $\mathbb{R}^{3}_{x,\cdots,x}=\mathbb{R}^{3}_{x}.$ The $n$ fold tensor product Hilbert space,
               \begin{equation}\label{bHc1nx}\bar{H}^{c}_{1,n,x,\cdots,x}=(\bigotimes_{j=1}^{n} \bar{H}^{c}_{j})_{x,\cdots,x} =(\bigotimes_{j=1}^{n}\bar{H}^{c}_{j})_{x}\end{equation} can be added to the fiber bundle for one and two particles with fibers at single points of $M$.

               The scaling function for this Hilbert space at $x$ is \begin{equation} \label{hy1yn} h(y_{1,n})_{x}=(\times_{j=1}^{n}g_{x}(y_{j}))^{1/n}= e^{\rho_{x}(y_{1,n})}\end{equation}where \begin{equation}\label{rhoxy1}\rho_{x}(y_{1,n})= \frac{\sum_{j=1}^{n}\gamma_{x}(y_{j})}{n}.\end{equation}Lifted into the fiber at $x,$ this scaling function has, as domain, $n$ tuples of points in $\mathbb{R}^{3}_{x}.$ The values are numbers of the form $e^{\rho(w_{1,n})_{h(y_{1,n})}}$ in the base set of $\bar{C}^{h(y_{1,n})}_{x}.$ Here $y_{1,n}=y_{1},\cdots,y_{n}$.  A similar equivalence holds for $w_{1,n}.$

               This shows that states for arbitrary  but fixed numbers of interacting particles can be described in fiber bundles with fibers associated with single points of $M$.  Additional components of quantum mechanics, such as momenta, energy and Hamiltonians can be included. This description is also suitable for relativistic quantum mechanics where states have components with different numbers of particles.

               \section{Conclusion}\label{Co}
               The description of the effect of number structure scaling on nonrelativistic quantum mechanics has been extended from that in earlier work.   In earlier work \cite{BenENSQM} the description was limited to states, momenta, and Hamiltonians for single particles.  Here states, momenta, and Hamiltonians for two or more interacting particles were described.  This includes the effects of scaling on entangled quantum states.

               For single particle states the effect of scaling  was to multiply the state amplitude at a point $y$ by the value of a complex scaling field, $g$ at $y$ where $g(y)=e^{\gamma(y)}.$  For $n$ particle entangled states the state amplitude at the $n$-tuple, $y_{1},\cdots,y_{n}$ of points was multiplied by a scaling factor that was the geometric average of the scaling factors for the individual points.  That is \begin{equation}\label{gy1n}g(y_{1},\cdots,y_{n})= (\times_{j=1}^{n}g(y_{j})^{1/n}=e^{\gamma(y_{1},\cdots,y_{n})}\end{equation}where
               \begin{equation}\label{gy1yn}\gamma(y_{1},\cdots,y_{n})=\frac{\sum_{j=1}^{n}\gamma(y_{j})}{n}.
               \end{equation}

               The mathematical framework of fiber bundles was used here to  describe the effects of number
               scaling. The fibers  of the bundles that include the description of $n$ interacting particles were associated with $n$-tuples of  space locations.  This led to the problem of how to combine bundle descriptions for different numbers of interacting particles into one fiber bundle picture.

               This problem was solved by noting that $n$-tuples of reference locations where the $n$ locations were all the same was equivalent to a description with each fiber at a single location.  The result was a single fiber bundle description for quantum mechanics of an arbitrary number of particles. For $n$ particles, the component of the fiber at $x$ consists of an $n$ fold tensor product of single particle Hilbert spaces.  The associated scaling factor is that of Eq. \ref{gy1n}.

               Much work remains.  This includes extension of the description to include relativistic quantum mechanics and general relativity.  Fiber bundles are well suited for this task. In addition one wants to determine the physical nature, if any, of the complex scaling field, $\gamma.$ The absence of any  experimental effect of the scaling field,  indicates that its local effect, must be below the sensitivity of experiments, at least so far. The effect of the field at cosmological distances is, at present, open.

             \section*{Acknowledgement}
        This material is based upon work supported by the U.S. Department of Energy, Office of Science, Office
        of Nuclear Physics, under contract number DE-AC02-06CH11357.

    \end{document}